\documentclass[sigconf,authorversion=true]{acmart}

\usepackage{balance}

\usepackage[utf8]{inputenc}
\usepackage[T1]{fontenc}

\usepackage{multirow}

\graphicspath{{incl/}}

\makeatletter
\newcommand*{\centerfloat}{%
	\parindent \z@
	\leftskip \z@ \@plus 1fil \@minus \textwidth
	\rightskip\leftskip
	\parfillskip \z@skip}
\makeatother

\begin{document}

% ISPD; do not delete this code.
\fancyhead{}

%\setlength{\textfloatsep}{9pt plus 2pt minus 4pt}
%\setlength{\floatsep}{6pt plus 2pt minus 2pt}
%\setlength{\dbltextfloatsep}{9pt plus 2pt minus 4pt}
%\setlength{\dblfloatsep}{6pt plus 2pt minus 2pt}

%\settopmatter{printacmref=false,printccs=false,printfolios=false}

\copyrightyear{2021}
\acmYear{2021}
\setcopyright{acmcopyright}\acmConference[ISPD '21]{Proceedings of the 2021 International Symposium on Physical Design}{March 22--24,
	2021}{Virtual Event, USA}
	\acmBooktitle{Proceedings of the 2021 International Symposium on Physical Design (ISPD '21), March 22--24, 2021, Virtual Event, USA}
	\acmPrice{15.00}
	\acmDOI{10.1145/3439706.3446902}
	\acmISBN{978-1-4503-8300-4/21/03}
%
% Submission ID. 
% Use this when submitting an article to a sponsored event. You'll receive a unique submission ID from the organizers
% of the event, and this ID should be used as the parameter to this command.
%\acmSubmissionID{123-A56-BU3}

% classification; code generated from
% http://dl.acm.org/ccs.cfm
%
\begin{CCSXML}
<ccs2012>
<concept>
<concept_id>10002978.10003001</concept_id>
<concept_desc>Security and privacy~Security in hardware</concept_desc>
<concept_significance>500</concept_significance>
</concept>
<concept>
<concept_id>10010583.10010786</concept_id>
<concept_desc>Hardware~Emerging technologies</concept_desc>
<concept_significance>500</concept_significance>
</concept>
</ccs2012>
\end{CCSXML}

\ccsdesc[500]{Security and privacy~Security in hardware}
\ccsdesc[500]{Hardware~Emerging technologies}

\title{%
	Hardware Security for and beyond CMOS Technology%: An Overview on Fundamentals, Applications, and Prospects
}
%\subtitle{%
%	An Overview on Fundamentals, Applications, and Challenges%, and Prospects
%}

\author{Johann Knechtel}
 \authornote{This work was supported in part by NYUAD REF under Grant RE218 and by the joint NYU/NYUAD Center for Cybersecurity (CCS).}
 \affiliation{Division of Engineering,
	 New York University Abu Dhabi, United Arab Emirates}
 \email{johann@nyu.edu}

\begin{abstract}
%% 132 words; limit is 200
%
As with most aspects of electronic systems and integrated circuits,
   hardware security has traditionally evolved around the dominant CMOS technology. However,
   with the rise of various emerging technologies, whose main purpose is to overcome the fundamental limitations for scaling and power consumption of
   CMOS technology,
   unique opportunities arise to advance the notion of hardware security.
In this paper, I first provide an overview on hardware security in general. Next, I review selected emerging
technologies,
namely (i)~spintronics, (ii)~memristors, (iii)~carbon nanotubes and related transistors,
	(iv)~nanowires and related transistors, and (v)~3D and 2.5D integration.
	I then discuss their application to advance hardware security and also outline related challenges.

\end{abstract}

\keywords{%
	Hardware Security; Spintronics; Memristors; Carbon Nanotubes (CNTs) Transistors; Nanowire Transistors; 3D Integration; 2.5D Integration;
	%Randomness; Variability; Polymorphic Behavior; Reconfigurability;
	Reverse Engineering; Tampering;
	%Counterfeiting;
	Theft of IP;
	%Intellectual Property;
	Hardware Trojans; Physical Attacks;
	%Probing Attacks;
%	Side-Channel Attacks; Fault-Injection Attacks;
	Data Security; Physically-Unclonable Functions (PUFs);
	True Random Number Generators (TRNGs); Logic Locking; Split Manufacturing; Camouflaging;
	%Masking; Monitoring;
	Root of Trust
}

\maketitle

\section{Introduction}
\label{sec:introduction}

% COMMENT general introduction; emerging devices, hardware security

In our modern age of omnipresent and highly interconnected information technology, (cyber)security becomes ever more challenged. Among many
other prominent incidents, e.g., in April 2019, more than 885 million records from First American Corporation, including bank account
details, Social Security digits, etc., were leaked publicly on the internet~\cite{krebs19}.
Within the realm of cybersecurity in general, hardware security in
particular is concerned about achieving security and trust directly within the underlying electronics. Therefore, hardware security seeks to
build up, e.g., so-called root of trust (RoT) schemes for isolation and attestation of computation~\cite{maene18,
	%zhang19,
	nabeel20_ROT_TC},
or other hardware primitives and protection schemes.

As with most aspects of modern electronic systems and integrated circuits (ICs), hardware security has traditionally evolved around the dominant
complementary metal-oxide-semiconductor (CMOS)
technology. However,
   with the rise of various emerging technologies, whose main purpose is to overcome the fundamental limitations for scaling and power consumption of
   CMOS technology,
   unique opportunities arise also to advance the notion of hardware security.
   For example, the use of memristive technology is promising for the secure management of secret keys~\cite{jiang18}, which is rather challenging for
   CMOS memory technologies.

% COMMENT scope, structure of this paper

In this paper, I review initially the fundamentals of hardware security and prior art for CMOS technology,
for both the key domains of (i)~data security at runtime and (ii)~confidentiality and integrity of hardware itself
(Sec.~\ref{sec:security}).
Then, I cover the fundamentals of
selected, prominent emerging technologies, namely (i)~spintronics, (ii)~memristors, (iii)~carbon nanotubes and related transistors,
	(iv)~nanowires and related transistors, and (v)~3D and 2.5D integration (Sec.~\ref{sec:emerging}).
Based on this review, promising matches of emerging technologies and their properties toward the needs for hardware security are
revealed (Sec.~\ref{sec:emerging_for_sec}).
Selected researched and (at least partially) demonstrated security schemes which are based on emerging technologies
are also discussed in Sec.~\ref{sec:emerging_for_sec}. In Sec.~\ref{sec:outlook},
I uncover challenges---and resulting chances---for future advancements, essentially advocating for further
interdisciplinary efforts, where the physical design community would want to become more involved as well.
Finally, I conclude in Sec.~\ref{sec:conclusion}.

\section{Fundamentals of Hardware Security and Prior Art for CMOS}
\label{sec:security}

Here, I review the fundamentals and prior art of hardware security in general. While it should be understood that this paper
can provide only an overview on this vast and fast-growing field, I strive to cover the most important aspects and seminal protection schemes.

\subsection{Data Security at Runtime}

The confidentiality, integrity, and available of data processing within electronics is subject to various threat scenarios, like
unauthorized access or modification of data,
	     side-channel and fault-injection attacks, and physical read-out and probing attacks.
Next, I provide an overview on these threats and related countermeasures.

\subsubsection{Unauthorized Access or Modification of Data}
Such ``bread-and-butter'' attacks are conducted mainly at the software level.
Cryptography represents the most commonly applied protection scheme, while
many hardware security features have been proposed and implemented as well, e.g., RoT architectures
\cite{maene18,
	%zhang19,
	nabeel20_ROT_TC} or
hardware crypto modules
%\cite{Secure_ICS_V10}
and true random number generators (TRNGs) to support the latter.
However, if not designed and implemented carefully, such security features become prone to hardware-centric attacks, e.g.,
see \cite{brier04,bayon16,qiu19}%
	%oflynn19,cui17}
	; such attacks are
discussed below.

\subsubsection{Side-Channel and Fault-Injection Attacks}
Side-channel attacks infer information from physical channels which are leaky due to sensitivities of electronics~\cite{zhou05}.
For example, it is well-known that the advanced encryption standard (AES) is vulnerable to power side-channel attacks when the hardware
implementation is unprotected \cite{brier04}.
%,oflynn19}.
For another example, modern processors leak information through caches and other buffers, related to timing behaviour, speculative execution, et cetera
\cite{osvik05,
	%lipp18,
	Schwarz2019ZombieLoad}.

Most countermeasures
%against side-channel attacks
apply some kind of ``hiding'' or {masking},
i.e., diffusion of the information leaked through side-channels, by various means taken 
from the system level \cite{GMK:16} down to gates \cite{bellizia18}.

Fault-injection attacks induce faults to aid deducing sensitive information.
Therefore, fault injection can also support or advance side-channels attacks.
Fault-injection attacks cover direct, invasive fault injection, e.g., by laser light \cite{selmke16} or electromagnetic waves
\cite{%cui17,
	bayon16,dehbaoui12},
as well as
indirect fault injection, e.g., by repetitive writing to particular memory locations \cite{vanderVeen16} or by deliberate ``misuse'' of dynamic voltage
and frequency scaling (DVFS) features~\cite{qiu19}.

Countermeasures include detection of faults at runtime, e.g., see \cite{NVKA:19}, and hardening against fault injection at design and
manufacturing time, e.g., see \cite{KGKP:18,
	%LM:06,
	DBC+:18}.
Note that
distinguishing between natural and malicious faults is non-trivial \cite{KGKP:18b}, which imposes practical challenges for recovery at runtime.

\subsubsection{Physical Read-Out and Probing Attacks}
An adversary with access to equipment used traditionally for failure analysis or inspection, like electro-optical probing or focused-ion beam
milling tools~\cite{principe17},
can mount quite powerful read-out attacks.
Among others, these attacks are: probing of transistors and wires~\cite{wang17_probing,helfmeier13},
either through the metal layers or the substrate backside;
monitoring of the photon emission induced by CMOS transistor switching~\cite{tajik17_CCS,krachenfels20_t-probing}; or
monitoring of electrical charges in memories~\cite{courbon16}.
When applied carefully, these attacks can reveal \textit{all} internal signals.

Countermeasures seek to prevent and/or detect the physical access.
Solutions include, e.g., shielding structures in the BEOL~\cite{
	%wang19_probing,
		lee19_shield,yi16}, deflection or scrambling structures 
in the substrate~\cite{shen18_ISTFA}, and detector circuitry~\cite{weiner18}.
Earlier studies such as~\cite{ishai03} also considered formally secure techniques. However, such schemes are subject to limitations
assumed for the attackers, which can become obsolete and would then render the formal guarantees void~\cite{krachenfels20_t-probing}.

\subsection{Confidentiality and Integrity of Hardware}

Besides the severe threats on data security at runtime, as outlined above, other threats such as reverse
engineering (RE), piracy of chip-design intellectual property (IP), illegal overproduction, counterfeiting, or insertion of hardware Trojans represent further 
challenges for hardware security.
These threats arise mainly due to the globalized and distributed nature of modern supply chains for electronics, 
which span across many entities and countries nowadays~\cite{rostami14}.

A multitude of protection schemes have been proposed, which can be broadly classified into
IP protection,
Trojan defense,
and physically-unclonable functions (PUFs).
All these schemes seek
to protect the hardware
from different attackers, which 
include untrusted foundries, untrusted testing facilities, untrusted end-users, or a combination thereof.
Next, I provide an overview on these schemes; more details can also be found in, e.g., \cite{knechtel19_IP_COINS,rostami14}.

\subsubsection{IP Protection}
This subject can be further classified, namely into
logic locking, camouflaging, and split manufacturing.
Camouflaging and split manufacturing alter the manufacturing process to protect against malicious end-users and untrusted foundries,
respectively, whereas logic locking works at the design level and seeks to protect against both, the foundries and the end-users.

\textit{Logic locking} protects the IP by inserting dedicated locks, 
which are operated by a secret key~\cite{YRS20}.
Without the secret key, logic locking ensures that
the details of the design IP cannot be fully recovered and
the IC remains non-functional.
The locks are commonly realized by additional logic, e.g., XOR/XNOR gates.
%	%~\cite{roy10},
%AND/OR gates
%	%~\cite{Dupuis2014novelLE},
%or look-up tables (LUTs)).
%	%~\cite{logicalbarriers}.
Only after manufacturing (preferably even only after testing~\cite{yasin16_test})
is the IC to be activated, namely by loading the secret key into a dedicated, tamper-proof on-chip memory.
%The realization of such
%memories remains under research and development, e.g., see \cite{
%	%tuyls06,
%anceau17}---this fact represents an obstacle towards wider application of logic locking.

Early works on logic locking considered various heuristics for insertion of
locks~\cite{roy10,
	%baumgarten2010preventing,JV-Tcomp-2013,Plaza_2015_TCAD,
	yasin16_locking_TCAD}. Upon dissemination of a powerful oracle-guided, \footnote{That is, a functional IC is required to obtain valid I/O observations.}
	Boolean satisfiability (SAT)-based attack \cite{subramanyan15}, however,
    %which defeated all known locking schemes at that time,
    the community had to develop advanced schemes as in \cite{xie16_SAT,yasin17_CCS} and others.
    In turn, these schemes stimulated the further development of other attacks and defense, e.g., \cite{shen17,shamsi17}, with some further
    considering also an
    oracle-less model, e.g., \cite{alrahis21_UNSAIL_TIFS,li19}.
    %,chakraborty18_SAIL}.

\textit{Camouflaging} serves to mitigate RE attacks
conducted by malicious end-users. Thus, camouflaging means to alter the layout-level appearance of an
IC in order to protect the design IP.
This can be achieved by dedicated front-end-of-line (FEOL) processing steps, like manipulation of dopant regions, gate structures, and/or gate 
contacts~\cite{rajendran13_camouflage,erbagci16,li16_camouflaging}, but also by obfuscation of the back-end-of-line (BEOL)
	interconnects~\cite{patnaik20_Camo_BEOL_TCAD}.
Camouflaging has been made available for commercial application, e.g., see the \textit{SypherMedia Library}~\cite{sypher}.
Note that obfuscation is also known in the context of design-time protection, e.g., by obfuscating finite state
machines~\cite{lao15}---such techniques are orthogonal to camouflaging.

As with logic locking, camouflaging is prone to analytical attacks~\cite{rajendran13_camouflage,yu17}.
In addition, camouflaging may be undermined by physical read-out and probing attacks outlined above.
See also~\cite{knechtel19_IP_COINS,vijayakumar16} for a more detailed overview on camouflaging.

\textit{Split manufacturing} seeks to protect the design 
IP from untrustworthy foundries~\cite{rajendran13_split, sengupta17_SM_ICCAD, sengupta19_LL_SM_DATE, patnaik18_SM_ASPDAC, patnaik18_SM_DAC, mccants16}.
As indicated by the term, the idea 
is to split up the manufacturing flow, most commonly
into an untrusted FEOL process and a subsequent, trusted BEOL process.
Such splitting into FEOL and BEOL is practical for multiple reasons: (i)~outsourcing the FEOL is desired, as it requires some
high-end and costly facilities, (ii)~BEOL fabrication on top of the FEOL is significantly less complex than FEOL fabrication itself,
(iii)~some in-house or trusted third-party facility can be engaged for BEOL fabrication,
	and
(iv)~the sole difference for the supply chain is the preparation and shipping of FEOL wafers to
that facility for BEOL fabrication.
Note that split manufacturing has been demonstrated successfully; \cite{vaidyanathan14_2}
describes promising results for a 130nm process split between \textit{IBM} and \textit{GlobalFoundries}, and \cite{mccants16}
reports on a 28nm split process run by \textit{Samsung} across Austin
and South Korea.

For the FEOL facility, a split layout appears as a ``sea of gates,'' making it
difficult for related adversaries to infer the entire netlist and its design IP.
Still, given that regular, security-agnostic design tools work holistically on both the FEOL and BEOL, hints on the omitted wiring can
well remain in the FEOL \cite{wang16_sm,li19_SM_ML_DAC,li20_SM_ML_TCAD}.

\subsubsection{Trojan Defense}
The notion of Trojans is wide-ranging and requires multiple dimensions for classification~\cite{BT18}---it relates to malicious hardware modifications that are (i) working at the system level,
register-transfer level (RTL), gate/transistor level, or the physical level; (ii) seeking to leak information from an IC, reduce the IC's performance,
or disrupt an IC's working altogether; (iii) are always on, triggered internally, or triggered externally; etc.
Trojans are likely introduced by
untrustworthy third-party IP,
adversarial designers, or
through ``hacking'' of design tools~\cite{basu19},
or, arguably even more likely, during distribution and deployment of ICs~\cite{Swierczynski2017}.\footnote{%
Although it has been projected traditionally as the main scenario, I argue that the likelihood of Trojans being introduced at fabrication
time is rather low.
That is because any such endeavour, once detected, would fatally disrupt the business of the affected foundry.
Therefore, foundries can be expected to employ all technical and organizational means available to them to hinder modifications by malign
employees.}

Defense schemes can be classified into (i) Trojan detection during design and manufacturing time
and (ii) Trojan mitigation at runtime.
The former relies on testing, verification, et cetera~\cite{
%CWPPB:09,
%AARP:10,
JM:08,LJM:12,guo19_QIFV,sugawara14,vashistha18,chandrasekharan15}, whereas
the latter
relies on dedicated security features for
testability and self-authentication~\cite{xiao14}, monitoring and detection of malicious
activities~\cite{kim11_trojan,bhunia13,basak17,wu16,wahby16}, etc.
Besides,
%IP protection schemes such as
logic locking and split manufacturing
can hinder Trojan insertion at manufacturing time, at least to some degree.
That is because an adversary without full understanding of the layout and its IP cannot easily insert some specific, targeted Trojans~\cite{imeson13}.

\subsubsection{Physically-Unclonable Functions (PUFs)}

When applied some input stimulus,
a PUF should provide a fully de-correlated output
response. This response should be reproducible for the very same PUF, even under varying environmental conditions, but it must differ 
across different PUF instances, even for the same PUF design.
PUFs are used for
(i)~``fingerprinting'' or authentication of hardware,
using so-called ``weak PUFs'' which provide capabilities for processing only one/few fixed
inputs,
or
(ii)~challenge-response-based security schemes, using so-called ``strong PUFs'' which provide capabilities for processing a
large number of inputs
\cite{herder14,maes10,chang17_PUF}.
Desired properties for PUFs are uniqueness, unclonability, unpredictability, reproducibility, and tamper resilience.

\textit{Electronic PUFs} represent the dominant class of PUFs, with prominent types of electronic PUFs using ring oscillators, arbiters, bistable rings, and
memories~\cite{maes10,herder14,ganji17_thesis,chang17_PUF}.
Such PUFs are relatively simple to implement and integrate,
even for advanced processing nodes.
The core principle for such PUFs is to leverage the process variations inherent to (CMOS) fabrication, through various
dedicated circuitry. 
However, the resulting randomness
is limited for most PUF implementations; it may be machine-learned and, thus, cloned~\cite{chang17_PUF,ruehrmair13,liu17,ganji17_thesis}.

\textit{Optical PUFs} represent another interesting class~\cite{pappu02,ruehrmair13_IACR,tuyls07,maes10,grubel17,knechtel19_peo-PUF_JLT}.
Here the idea is to
manufacture an ``optical token'' which, in addition to structural variations inherently present in selected optical media, 
may contain randomly included materials, e.g., nanoparticles.
%Besides such a token, optical PUFs require further components, for generating the optical input and processing the output.
%The fundamental phenomena underlying an optical PUF are scattering, reflection, coupling, and absorption of light within the optical token.
Depending on the materials used for the token and the inclusions as well as the design of the token itself, these phenomena can be highly
chaotic and nonlinear by nature~\cite{knechtel19_peo-PUF_JLT,grubel17}.
Hence, optical PUFs are considered more powerful than electronic PUFs.

\section{Fundamentals of Selected Emerging Technologies}
\label{sec:emerging}

%Next, I review emerging technologies.
It should be understood that
I can provide only an overview on selected emerging technologies; there are further technologies, like negative capacitance
field-effect transistors
(NCFETs)~\cite{amrouch19,knechtel20_PSC_Micro} or photonics~\cite{perez17,orcutt12}.
%, which are not covered in this paper.
Aside from 3D and 2.5D integration, note that all selected technologies are realized at the device level;
therefore, these technologies can also be referred to as \textit{emerging devices.}

The selected technologies are all compatible with CMOS manufacturing, at least to some degree. Thus,
these technologies appear promising and practical for the near future,
as they can also realize some hybrid CMOS-emerging electronics.
In general, emerging technologies seek to overcome fundamental limitations for CMOS regarding scalability and power consumption, among other
aspects.
In practice, various technologies are also applied in conjunction, e.g., the \textit{N3XT} approach
by \textit{Stanford University}
researchers
and others
leverages carbon nanotubes and spintronics within 3D ICs~\cite{aly19}, and Wang and Chen studied spintronic memristor devices~\cite{wang10_mr}.

\subsection{Emerging Devices}

\subsubsection{Spintronics}

Also known as \textit{spin electronics}, spintronics differ from CMOS technology in various
aspects~\cite{baek18,manipatruni18,nikonov13,makarov16,fong16,rangarajan19_GSHE_DT}.
First and foremost, in addition to an electronic charge,
the \textit{spin} of electrons is leveraged for both computation and storage/memory.
Second, the switching process is non-volatile, magnetoelectric, and subject to other related phenomena like spin-transfer torque (STT).
Third, spintronics are implemented typically as stack of heavy metals, ferromagnets, and/or oxide
structures~\cite{baek18,manipatruni18,nikonov13,makarov16,fong16,rangarajan19_GSHE_DT},
but the use of other materials has been proposed as well, e.g., graphene~\cite{Han2014}, superconducting materials~\cite{Linder2015}, or even organic
materials~\cite{Rocha2005}.
Still, manufacturing of spintronics can be made compatible with CMOS processing~\cite{makarov16,manipatruni18}.
Fourth, in comparison to CMOS, spintronic devices can offer lower power consumption,
built-in memory functionality,
built-in reconfigurability,
and better scalability~\cite{makarov16,fong16,nikonov13,manipatruni18}.

Spintronics have been studied in detail for memory~\cite{makarov16,fong16,BHATTI2017530} and/or
logic~\cite{nikonov13,manipatruni18,baek18,rangarajan17_TMAG,rangarajan19_GSHE_DT,makarov16,fong16} applications,
and even for interconnects~\cite{naeemi14}.
For example, efforts led jointly by \textit{Intel}, \textit{UC Berkeley}, and \textit{Berkeley Lab} promote a type of magnetoelectric spin-orbit logic that
has superior switching energy (by a factor of 10 to 30), lower switching voltage (by a factor of 5), and enhanced
logic density (by a factor of 5) when compared to CMOS.
This magnetoelectric device is also compatible with CMOS manufacturing, as it can be implemented in the interconnect layer.
Among other applications, reconfigurable logic, probabilistic computing, and in-memory computing are good matches for
spintronics~\cite{makarov16,matsunaga08,rangarajan17_TMAG}.

\subsubsection{Memristors}

The memristor, short for \textit{memory resistor}, represents another fundamental circuit element besides the well-known resistor, capacitor, and
inductor elements; its theory was studied already in 1971~\cite{chua71}.
Memristors retain an internal resistive state according to the history of voltage or current applied.
Another interesting characteristic for some but not all memristors is a nonlinear response, resulting in ``pinched hysteresis loops.''
That is, such memristors exhibit a current/voltage threshold, with their internal state only being switched when this threshold is exceeded.

The implementation of memristor devices remains under research and development, considering various materials and arrangements like
titanium dioxide~\cite{Torrezan_2011}, spintronics~\cite{wang10_mr}, or carbon nanotubes~\cite{ILINA2017514}, with most approaches
remaining compatible with CMOS fabrication.
Memristive systems in the broader sense, like resistive random-access memories (ReRAMs) or even phase-change memories (PCMs),
%\footnote{While some
%argue that such related technologies should be considered as memristive ones as well~\cite{Chua2011,6518267}, others argue against that and for a more strict
%interpretation of memristive properties in general~\cite{Di_Ventra_2013,meuffels2012fundamental}.}
are progressing towards commercial application~\cite{7049528,7046994}.
Aside from memory, memristors are also interesting for in-memory computing, neuromorphic computing, and reconfigurable
logic~\cite{6518267,tetzlaff2013memristors,cai19}.

\subsubsection{Carbon Nanotubes and Transistors}

% CNTs in general
Carbon nanotubes (CNTs) comprise one or more rolled-up layers of \textit{graphene}, the planar arrangement of single-layer carbon atoms in 2D honeycomb-like
structures. In other words, CNTs form cylindrical structures with single or multiple ``walls'' made of carbon sheets.
CNTs are typically few nanometers in diameter and few micrometers in length.
CNTs can be either metallic conductors or semiconductors, depending on their structure.
CNTs exhibit outstanding electrical, physical, and thermal properties~\cite{CNT_TSDM17,Anantram_2006,lienig2018mitigating}, mainly due to the strong bonds between their carbon atoms.
For example, individual metallic CNTs can, in principle, hold current densities more than 1,000 times greater than copper, which is also
promising to mitigate electromigration~\cite{lienig2018mitigating}.
In practice, however, CNTs have to form interfaces with each other and with other materials~\cite{nl203701g,CNT_TSDM17,Anantram_2006,8454842},
pushing such gains somewhat out of reach.
Still, one can also build up composite structures to tune CNT properties as needed, e.g., using copper
to adapt the thermal expansion coefficient of CNTs toward that of silicon~\cite{C3NR05290G}.

% CNTs for interconnects
CNTs have been studied extensively for interconnects, e.g., see \cite{CNT_TSDM17,8454842,Anantram_2006}, as well as
for transistors, e.g., see \cite{shulaker13,aly19,wu18,nl203701g}.
%
% CNTs for FETs
In essence, carbon nanotube field-effect transistors (CNTFETs or CNFETs) leverage multiple CNTs as transistor channels, which can be realized in
different arrangements, e.g., as gate-all-around structure~\cite{4435964}.
CNTFETs are subject to the imperfection and variability of CNT manufacturing. However, these limitations can be addressed by device and
chip-design methodologies~\cite{zhang12_CNT}---chip-scale application of CNTFETs has been demonstrated successfully~\cite{shulaker13,wu18}.

\subsubsection{Nanowires and Transistors}

Nanowire FETs (NWFETs) leverage nano-scaled and semiconductive wires as transistor channel. Various types of NWFETs have been
studied, e.g., using silicon~\cite{mikolajick17}, indium arsenide~\cite{1626445}, germanium, or even polymers~\cite{BRISENO200838} for materials;
homogeneous or heterogeneous wire structures~\cite{4668552}; gate-all-around~\cite{COLINGE201133} or vertical gate structures~\cite{1626445};
et cetera---an overview can also be found in \cite{4668552}.
Conceptionally, NWFETs are somewhat similar to CNTFETs, but NWFETs allow for finer control of desired properties during manufacturing (albeit
challenges for chip-scale manufacturing are there as well~\cite{4668552}), whereas CNTFETs offer better performance~\cite{Singh2016}.
Besides, NWFETs are somewhat similar to nanosheet transistors, which are progressing already towards commercial application~\cite{HOOK20181}.

Nanowire transistors have been proposed for sensing applications~\cite{VU2010354}, flexible electronics~\cite{4668552},
and reconfigurable logic~\cite{mikolajick17}, among other applications.
For \cite{mikolajick17}, an additional program gate enables to switch the transistor between n-channel and p-channel behavior, by
selectively suppressing the injection of one type of charge carriers (e.g., electrons), while the other type (e.g., holes) is modulated via the control gate.

\subsection{3D and 2.5D Integration}

Aside from the emerging devices outlined above,
      which are all realized at the device level,
3D and 2.5D integration targets at the system
level. That is, these technologies embrace notions of ``building skyscrapers'' or ``city clusters'' of
electronics~\cite{knechtel17_TSLDM,radojcic17,EF16,iyer15}.
Two factors drive these technologies:
for one, that is the CMOS scalability bottleneck, which becomes more exacerbated for advanced nodes by issues like routability, pitch scaling,
and process variations,
for another, that is the need to advance heterogeneous and system-level integration.
Both drivers are also known as ``More Moore'' and ``More than Moore,'' respectively.

\textit{3D integration} means to vertically stack and interconnect multiple chips or active layers.
This approach can be classified by the underlying technology, with the main ones being
	(i) through-silicon via (TSV)-based 3D ICs,
	 where multiple chips are fabricated separately and then stacked and bonded,
	 %The vertical
	%interconnects across the 3D ICs are realized by relatively large metal TSVs which are cutting through the individual chips.
	(ii) face-to-face (F2F) 3D ICs,
	where two chips are fabricated separately and then bonded directly at their BEOL metal ``faces,''
	%TSVs or wirebonds
	%are commonly used for external connections.
	and (iii) monolithic 3D (M3D) ICs~\cite{knechtel17_TSLDM},
	where multiple active layers are manufactured sequentially.
	%The vertical interconnects are implemented by
	%regular metal vias.
Various studies, prototypes, and commercial products have shown that such 3D ICs offer significant benefits over 2D
ICs, e.g., see \cite{fick13, iyer15, kim12_3dmaps, shilov18}.

\textit{2.5D integration}, otherwise known as \textit{interposer technology}, facilitates system-level integration of
2D/3D ICs in side-by-side fashion.
That is, an interposer serves as integration carrier, accommodates some system-level interconnect
fabric, and may even contain active components~\cite{
		%COMMENT redudant refs, also for active interposer below
		%lau11,clermidy16,
	pavlidis17,manoj17,kim19,
%COMMENT refs for active interposer
stow17,clermidy16,takaya13,lau11,vivet20}.
Building advanced electronic systems in the form of 2.5D ICs is considered less complex than 3D
integration~\cite{stow17,pavlidis17,lau11}. 
That is also because interposers are typically implemented using mature technology nodes, for cost savings and yield management.
Finally, 2.5D ICs are already well-established in the market, e.g., see \cite{lee16,shilov18}.

\section{Emerging Technologies for Hardware Security}
\label{sec:emerging_for_sec}

\begin{figure*}[tb]
\centerfloat
	\includegraphics[width=1.08\textwidth]{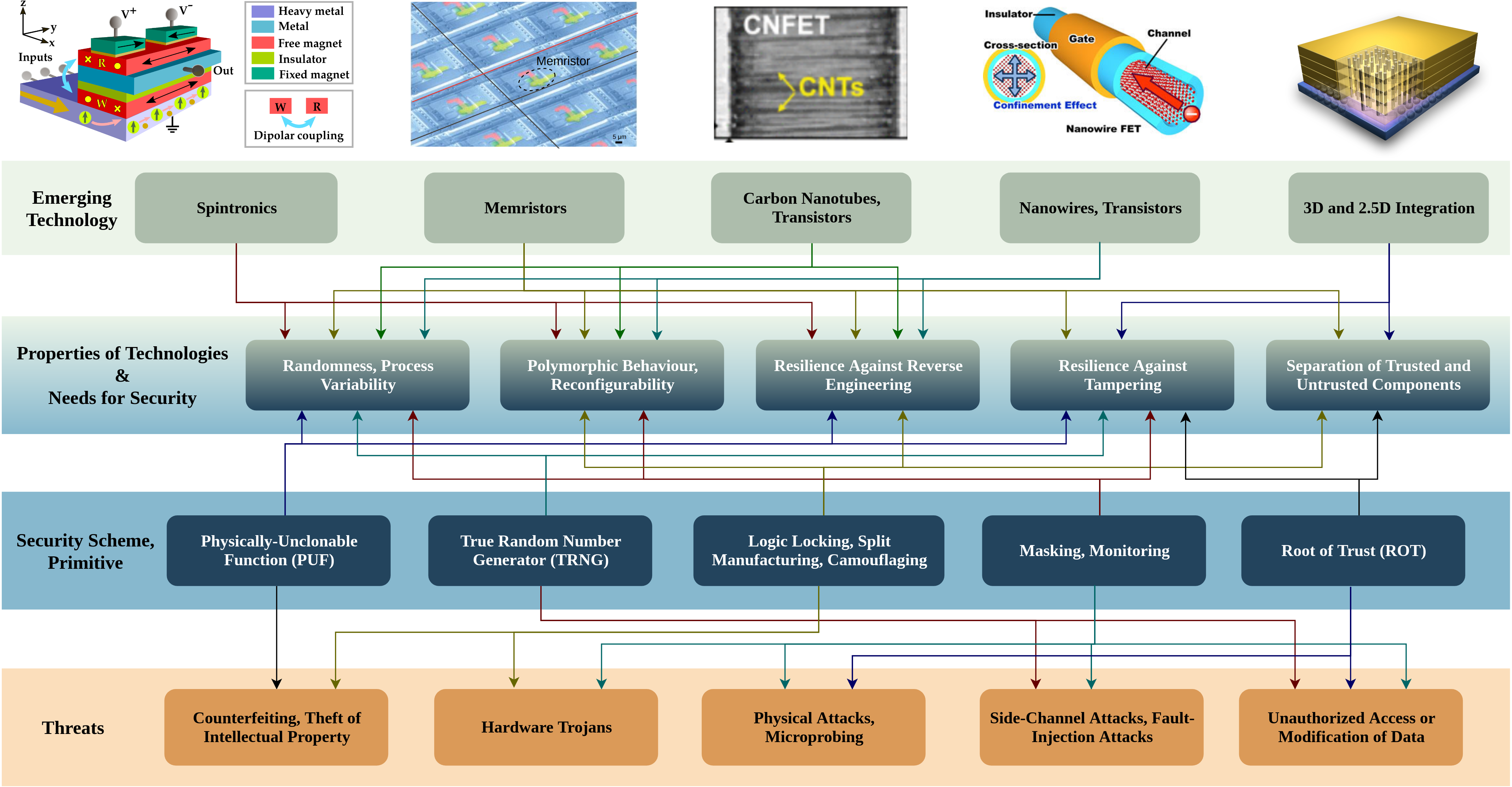} 
	%\smallerspacecaption
	\caption{A selective overview on emerging technologies, their properties, matching security schemes, and countered threats.
	}
	\label{fig:overview}
\end{figure*}

Various emerging technologies offer the potential to advance the notion of hardware security.
In Fig.~\ref{fig:overview}, I outline the selected emerging technologies covered in this paper,
   their properties relevant and beneficial for hardware security, the
security schemes which are supported accordingly, and the security threats countered by such schemes. While this illustration is certainly not
an overarching one, it does consider all the aspects introduced in this paper.

\subsection{Emerging Devices}

The reviewed emerging devices have some interesting properties in common, which are more difficult to achieve in traditional CMOS technology.
More specifically, spintronics, memristors, CNTFETs, and NWFETs can all be tailored to comprise significant variability/randomness,
reconfigurability or polymorphic behavior, resilience against reverse engineering,
		  and possibly also for separation of trusted and untrusted parts (the latter by means of split manufacturing).
Therefore, these devices can serve well for PUFs, TRNGs, IP protection schemes, and to mask side-channel leakage.
Moreover, memristors may also offer resilience against tampering, by means of destructive data
management~\cite{jiang18}.

It should be understood that the prospects for actual implementation of such security schemes based on emerging devices depends on various
aspects, ranging from circuit design and security analysis in general, down to manufacturing capabilities and device maturity, among others.
Still, there is a wide range of prior art developing and promoting such schemes, of which some are reviewed next.
Note that other papers have reviewed emerging devices in the context of hardware security 
as well, e.g., see~\cite{rajendran15_nano,ghosh16,rahman17}.

\subsubsection{Spintronics}

%\cite{parveen17, yang18, patnaik18_GSHE_DATE, patnaik19_GSHE_TCAD, rangarajan20_MESO_TETC} polymorphic behavior, reconfigurability; obfuscation, camouflaging, locking
%
Various studies proposed polymorphic behavior and/or reconfigurability for IP protection.
%obfuscation, camouflaging, and/or logic locking.
%
For example, Alasad~\textit{et al.}~\cite{alasad2017leveraging} use all-spin logic
for camouflaging. However, the layouts for some of their proposed device primitives are unique;
they can be readily distinguished during imaging-based reverse engineering.
Besides, their primitives suffer from relatively high energy consumption, with around 350 $\mu$W consumed for ns-range delays.
In \cite{winograd2016hybrid,yang18}, the authors introduced spintronics-based reconfigurable lookup tables (LUTs) for design obfuscation.
However, these approaches can fall short regarding their resilience against SAT attacks~\cite{patnaik19_GSHE_TCAD}.
Also note that such approaches are conceptionally similar to design obfuscation leveraging traditional field-programmable gate arrays (FPGAs).
In \cite{parveen17} and \cite{patnaik18_GSHE_DATE, patnaik19_GSHE_TCAD},
polymorphic and obfuscated logic has been studied, based on domain-wall motion devices and
on giant spin-Hall effect (GSHE) devices, respectively. One important benefit of the latter studies over the former
is that they support all 16 possible functions per device; this renders these devices superior in terms of SAT
resilience~\cite{patnaik18_GSHE_DATE, patnaik19_GSHE_TCAD}.

In \cite{rangarajan20_MESO_TETC}, the notion of ``dynamic camouflaging'' based on polymorphic magnetoelectric spin-orbit (MESO) devices has
been introduced. Unlike regular camouflaging, this notion can also protect against adversaries in the foundries and test facilities, as the true
functionality is configured only later on within the polymorphic fabric. Thus, this notion is also conceptionally similar to logic locking;
however, unlike with locking, no additional devices or gates are required to realize this kind of security.

It has been noted that spintronics can offer some resilience against side-channel
attacks~\cite{patnaik19_GSHE_TCAD,rangarajan20_MESO_TETC,roohi19}. For example, the
magnetoelectric switching of these devices does not emit photons; related attacks as in~\cite{tajik17_CCS} can be ruled out to begin with.
With spintronics used for logic, fault-injection and side-channel attacks based on magnetic fields or temperature curves may be more difficult to
achieve~\cite{patnaik19_GSHE_TCAD,rangarajan20_MESO_TETC}, unlike with spintronics used for memories~\cite{ghosh16}.
Moreover, in \cite{roohi19} the authors used spintronics to build up polymorphic circuitry and different circuit templates
for the same functionality, which are randomly switched between at runtime, in order to mask the power side-channel.

In \cite{iyengar15S,ghosh16}, the authors advocate the process variations for nanowire manufacturing in domain-wall memories for PUFs.
In \cite{rangarajan17_TRNG}, the authors
leverage the inherently stochastic spin switching mechanism in nanomagnets
for TRNGs. Through device-level simulations, the authors demonstrate that their TRNG device works across large temperature
ranges, is immune to process variations, and can be implemented with significantly smaller layout costs than CMOS TRNGs.
In \cite{rangarajan20_SMART_ACC}, the authors propose an antiferromagnets-based secure memory scheme, which offers protection against
tampering, side-channel, and read-out attacks, and also promises orders of lower energy-per-bit than STT-RAM or PCMs.

Most studies focus on circuit design and security analysis, but few on technology aspects. While spintronics
are making fast progress toward application,
    it seems important to consider technology exploration also within the concerned security studies.

\subsubsection{Memristors}

The potential of using memristors for hardware-security schemes has been recognized already some years ago, e.g.,
for PUFs leveraging the process variations and the stochastic operation of memristors in 2013~\cite{rose13}.
More recently, another PUF concept has been proposed~\cite{nili18}, which leverages the non-linear I-V characteristics of memristors
(``pinched hysteresis'') and
applies analogue tuning of the memristor conductance, to increase the performance
and practicality of such PUFs and to reduce the complexity of the
peripheral circuitry. The authors of~\cite{nili18} provided an experimental demonstration and measurement results for their PUF concept.

In \cite{jiang18}, a memristive crossbar array is at the heart for secure management of secret keys.
The authors propose to combine unique fingerprints of memristor devices with storage of key values within those devices.
They construct control circuitry such that upon extraction of the fingerprints (for verification of authenticity of the chip), the key is destroyed.
Therefore, the secret key remains ``alive'' on the chip to enable its functionality (following the notion of logic locking) until any read-out is
conducted.
The authors of \cite{jiang18} provide an experimental demonstration and measurement results for their concept.
Such a concept is an important step for the practicality of logic locking, which requires tamper-proof memories for its security
promises against malicious end-users in the field.

In \cite{rezaei19}, the authors propose polymorphic circuitry for obfuscation also in the context of memristors. That is
possible because, in principle, the functionality of memristor devices within such obfuscated logic can be reconfigured. While the
authors provide a first study at the circuit and layout level---albeit without details on the technology exploration and library characterization---they do not provide any experimental demonstration.
Moreover, other studies caution about delay and power consumption for memristor-based logic unless circuit structures are
optimized~\cite{7811204}, which seems to conflict with obfuscation principles.

\subsubsection{CNTs and CNTFETs}

In \cite{liu18}, the authors propose PUFs which leverage the manufacturing variability of CNTs along with the notion of Lorenz chaotic
systems. The latter serves to enhance the decorrelation of inputs and outputs for such PUFs and, thus, renders them more resilient against
machine-learning attacks.
%, as the authors demonstrate.

In \cite{suresh16}, the authors conduct a simulations-based study on CNTFETS concerning Trojan detection, power side-channel leakage, and
camouflaging, and they find that CNTFETs are more promising in all aspects when compared to the traditional CMOS technology.

In \cite{rahman17}, the authors review the use of CNTS for PUFs, TRNGs, and also propose the technology to be used for novel sensors detecting
microprobing or other invasive attacks.

\subsubsection{Nanowires and
	NWFETs}

In \cite{bi16_JETC}, the authors propose silicon NWFETs for camouflaging. More specifically, they leverage the controllable ambipolarity
in NWFETs to build up a camouflaging primitive comprising the NAND, NOR, XOR, and XNOR functionalities. The authors also build up a polymorphic
NAND/NOR gate, and they present circuit-simulation results. In \cite{patnaik19_GSHE_TCAD}, however, it was shown that such primitives are prone to SAT attacks.

%In \cite{mikolajick17}, the authors indicate that their concept of reconfigurable NWFETs can also be used for hardware security, among other applications, but they do not provide
%further details.
In \cite{rai20_RFET_TETC}, the authors
first explore how transistor-level reconfigurability can be leveraged for logic locking and
split manufacturing in the context of silicon NWFETS models. Second, they
study how the very reconfigurability can be exploited to induce either short-circuit currents or
open-circuit configurations, essentially annihilating the reliability and functional properties of the chip; the authors argue that this critical property of
reconfigurable NWFETs could be either maliciously exploited as reliability-centric Trojan or intentionally leveraged as ``kill switch.''

In \cite{cui14_plasmonics, park16_plasmonics} the use of nanowires with plasmonics interaction upon optical inspection is proposed and
experimentally demonstrated. This idea serves for labelling and authentication of chips (or other goods, for that matter). Loosely related,
because without the need for nanowires, the authors in \cite{knechtel19_peo-PUF_JLT} proposed the concept of plasmonics-enhanced optical PUFs and
provide physical-simulation results and a security analysis.

\subsection{3D and 2.5D Integration}

The main benefits provided by 3D and 2.5D integration to advance hardware security are
(i)~the \textit{physical separation} of components,
be it across interconnects, active devices, or both, and
(ii)~the \textit{physical enclosure} of components, to shield them from
adversarial activities in the field.
In Table~\ref{tab:prior_3D_2.5D}, I summarize selected
works; I discuss these and others in some detail next.
Note that other papers have reviewed the benefits and fallacies for hardware security arising by 3D and 2.5D
integration as well, e.g., see~\cite{knechtel19_3D_IOLTS,xie16,dofe17}.

\begin{table}[tb]
\centering
\footnotesize
\caption{Selected Works Leveraging 2.5D/3D Integration for the Benefit of Hardware Security}
\label{tab:prior_3D_2.5D}
\setlength{\tabcolsep}{2.5mm}
\begin{tabular}{cccccc}
\hline
\textbf{Reference} & 
\textbf{Style} & 
\textbf{Scope; Means} & 
\textbf{Trusted Part} \\ 
\hline 
\multirow{2}{*}{\cite{valamehr13}} & \multirow{2}{*}{TSV} & Runtime monitoring; & \multirow{2}{*}{Whole 3D IC} \\
& & split manufacturing (SM) & \\ \hline
\cite{xie17} & 2.5D & IP protection; SM & Interposer \\ \hline
\cite{imeson13} & 2.5D & Trojan prevention; SM & Interposer \\ \hline
\cite{gu18} & F2F & IP protection; SM, camouflaging & Parts of 3D IC \\ \hline
%NOTE probably not only SM; check
%\cite{tezzaron08} & 3D & SM & IP Piracy & Gates \& Wires & FEOL \& BEOL \\ \hline
\cite{yan17_camo} & M3D & IP protection; camouflaging & Whole 3D IC \\ \hline
\multirow{2}{*}{\cite{patnaik19_3D_TETC}} & \multirow{2}{*}{F2F} & IP protection, Trojan prevention; & \multirow{2}{*}{Only BEOL} \\
& & SM, camouflaging & \\ \hline
\cite{cioranesco14} & TSV & Probing protection; enclosure & Whole 3D IC \\ \hline
\cite{knechtel17_TSC_DAC,bao17} & TSV & Side-channel mitigation; enclosure & Whole 3D IC \\ \hline
\cite{nabeel20_ROT_TC} & 2.5D & Runtime monitoring; separation & Interposer \\ \hline
%COMMENT yan17_camo supersedes dofe16_mono3D
%%~\cite{dofe16_mono3D} & Mono 3D & IP Piracy & Gates & BEOL? \\ \hline
%
\end{tabular}
\end{table}

\subsubsection{Confidentiality and Integrity of Hardware:
	Logic Locking}
To the best of our knowledge, 3D and 2.5D integration has not been explored yet for logic locking.
In the loosely related work
\cite{sengupta19_LL_SM_DATE}, the authors
leverage locking principles
to advance the notion of split manufacturing.
More specifically, they
lock the FEOL and delegate the unlocking to a separate, trusted BEOL facility.
The authors note that their scheme can also be unlocked at the package or board level, which might well suggest an
implementation as 2.5D IC.

\subsubsection{Confidentiality and Integrity of Hardware:
	Camouflaging}
The authors of~\cite{yan17_camo} were the first to propose camouflaging dedicatedly for 3D integration,
more specifically for M3D ICs.
The authors developed and characterized custom M3D camouflaged libraries, and they evaluated their scheme
at the gate level and at chip scale.
%	, considering a SIMON block cipher implementation and other benchmarks circuits.

The camouflaging is realized by dummy contacts, which has been proposed previously for camouflaging in classical 2D ICs.
Thus, while conceptionally not new, the work in~\cite{yan17_camo} leverages the benefits provided by M3D ICs in an effort
to advance the scalability
%and applicability
of camouflaging.
That is noteworthy because prior art for camouflaging
may
incur considerable layout cost.
For example, the NAND-NOR-XOR primitive of~\cite{rajendran13_camouflage} would incur 5.5$\times$ power, 1.6$\times$ delay, and 4$\times$ area cost
compared to a regular NAND gate. In practice, such cost allow for only few gates being camouflaged;
in turn, limited camouflaging scales
renders such schemes prone to SAT attacks~\cite{yu17,shamsi17}.
%and/or requires dedicated SAT-resilient schemes which, however, cannot prevent the theft of an
%approximate version of the IP~\cite{shamsi17}.}
In contrast, the work in~\cite{yan17_camo} report on average only 25\% power cost, 15\% delay cost, and 43\% area savings compared to regular
2D gates.

\subsubsection{Confidentiality and Integrity of Hardware:
	Split Manufacturing}
To advance split manufacturing via 3D and 2.5D integration seems both straightforward and promising.
That is because 3D and 2.5D integration allows to split a design into multiple chips, which can maintain their FEOL and BEOL layers independently
as is, whereas the overall 2.5D/3D stack can comprise further parts of the system-level interconnects.
Moreover, concerns regarding the practicability of classical split manufacturing---which are still prevalent, despite proof-of-concept studies
like~\cite{vaidyanathan14_2,mccants16}---can be elevated due to this very fact that individual chip would not have to be split manufactured, but
only the overall system.

The idea of such ``3D split manufacturing'' was outlined in 2008, by \textit{Tezzaron Semiconductor}~\cite{tezzaron08}.
Various studies are hinting at 3D split manufacturing as well, but most have some limitations. For example, the study~\cite{dofe17}
%% dofe17 supersedes gu16
%and \cite{gu16}
remains only on the conceptional level, while the studies~\cite{xie17,imeson13} utilize 2.5D integration with
``only'' wires being hidden from untrusted facilities. The latter is in principal equivalent to traditional split manufacturing but seems more
practical; still, the studies~\cite{xie17,imeson13} report on considerable layout cost.
Later on, \cite{devale17, gu18, patnaik18_3D_ICCAD, patnaik19_3D_TETC} promoted ``native 3D split manufacturing,'' i.e., with logic being split
across trusted and untrusted facilities.

%COMMENT could be dropped if space needed
One important finding of those later studies~\cite{devale17,gu18,patnaik18_3D_ICCAD,patnaik19_3D_TETC} is that the 3D partitioning as well as the vertical interconnect fabric both play an important role
and define a cost-security trade-off as follows: the more the design is split up across multiple chips, the higher the layout cost, 
due to the need for more vertical interconnect links and related circuitry,
but the more flexible and easier it becomes to ``dissolve'' the IP across the 3D stack.

Note that \cite{gu18,patnaik18_3D_ICCAD,patnaik19_3D_TETC} proposed 3D split manufacturing in conjunction with camouflaging.  While the
study~\cite{gu18} applies regular, FEOL-centric camouflaging, the studies~\cite{patnaik18_3D_ICCAD,patnaik19_3D_TETC}~argue that another
camouflaging approach is more appropriate for 3D split manufacturing, namely the obfuscation of the vertical interconnects.
Other works also suggest camouflaging at the system level.
For example, \cite{dofe16} proposed to obfuscate the vertical interconnect fabric of 3D ICs
by rerouting within dedicated network-on-chip (NoC) chips ``sandwiched'' between the regular chips.
This idea is conceptionally similar to the notion of randomized routing in~\cite{patnaik18_3D_ICCAD,patnaik19_3D_TETC},
but more flexible, yet also more costly---it seems only warranted in case 3D NoCs are to be employed in any case.

\subsubsection{Confidentiality and Integrity of Hardware:
	Trojan Defense}

In~\cite{patnaik19_3D_TETC}, the authors leveraged the benefits provided by 3D split manufacturing to advance the formally-secure but high-cost
	scheme of~\cite{imeson13} to mitigate Trojan insertion at manufacturing time.

Besides, 3D and 2.5D ICs seem rather
more vulnerable than 2D ICs to Trojan insertion during design and manufacturing time.
For example, the study in~\cite{mossa17} considered the negative bias temperature instability (NBTI) effect as stealthy Trojan trigger,
    motivated by the fact that thermal management is a well-known challenge for 3D ICs.
In a more general manner, I caution that the broader landscape of suppliers and actors involved with 3D and 2.5D integration can
open up new opportunities for attackers to embed Trojans.
Such a concern has also been voiced recently in~\cite{bunnie_36c3}, along with the wide-spread adoption of wafer-level chip-scale packaging (WLCSP).
The hypothesized attack here is that some malicious integration 
facility could place a thin Trojan chip between the target chip and the package microbumps, while that Trojan chip would contain TSVs to
both pass-through and tap into those external connections, gaining access to all these signals at will.
To mitigate detection by visual or X-ray inspection, it was argued that aligning those TSVs with the microbump locations might suffice.

Trojan detection at runtime, however, can benefit from 3D and 2.5D integration. That is because related security features can be implemented separately
using a trusted fabrication process and integrated/stacked later on with the commodity chip(s) to be
monitored~\cite{wahby16,nabeel20_ROT_TC}; see also the discussion on data security below.
	
\subsubsection{Confidentiality and Integrity of Hardware:
	PUFs}
The integration of multiple chips into 3D/2.5D stacks seems beneficial for the notion of PUFs, as the individual chips are subject to independent process
	variations. Thus, one can build up PUFs using multiple, independent sources of entropy.
In~\cite{wang14_SuperPUF, wang15_TSV-PUF}, two such schemes have been proposed, which further leverage the process variations of TSVs.
While promising in principle, these studies did not consider state-of-the-art machine learning attacks
like~\cite{chang17_PUF,ruehrmair13,liu17,ganji17_thesis}; their actual resilience remains yet to be demonstrated.

\subsubsection{Data Security at Runtime:
	Unauthorized Access or Modification of Data}

3D and 2.5D integration enables physical separation of components and, thus, allows for
trustworthy realization of security features like runtime monitors~\cite{mysore06,valamehr13,nabeel20_ROT_TC} or verifiers~\cite{wahby16}.

Still, I caution that the physical implementation of such schemes may become a
vulnerability by itself.
In~\cite{valamehr13}, for example, the authors propose introspective interfaces which, however, require additional logic within the commodity
chip to be monitored. It is easy to see that these interfaces would fail once they are modified by some malicious actor(s) involved with the design
or manufacturing of that commodity chip.  Thus, an undesirable dependency arises, possibly thwarting the scheme altogether. Note that the
authors themselves acknowledge this limitation in~\cite{valamehr13}.

In~\cite{nabeel20_ROT_TC}, a 2.5D root of trust has been proposed, which integrates
untrusted commodity chips/chiplets onto an active interposer which contains security features and further forms the backbone for system-level communication
between the chips/ciplets. Thus, a clear physical separation into commodity and security components exists, avoiding any 
security-undermining dependencies.

\subsubsection{Data Security at Runtime:
	Side-Channel and Fault-Injection Attacks}

In general, side-channel attacks seem to become more difficult for 3D and 2.5D ICs, considering the higher density of active devices and the more
complex circuit structures and architectures, which can result into more noisy side-channels.
For example, the authors of \cite{dofe18_PDN} studied power side-channel attacks on 3D ICs, and they observed that the power noise profiles from
the different chips within the 3D IC are superposed.
They also propose a randomized cross-linking scheme of voltage supplies for cryptographic modules, to render attacks on such modules more difficult.

Some prior art also studied side-channel attacks targeted explicitly for 3D ICs.
For example, \cite{gu16_ICCD} and \cite{knechtel17_TSC_DAC} demonstrate that thermal side-channel attacks on 3D ICs can be mitigated
at runtime and at design time, respectively.
However, the approach in \cite{gu16_ICCD} seems less practical; to mitigate information leakage through thermal patterns,
    it leverages dynamic generation of additional dummy activities,
which exacerbates the challenge of thermal management for 3D ICs even further.
In contrast, the authors of \cite{knechtel17_TSC_DAC} model the impact of TSVs and module placement on heat distribution
as well as thermal leakage during floorplanning, thereby enabling leakage mitigation along with reduction of peak temperatures.

Besides, some studies leverage 3D
and 2.5D integration to advocate for security schemes otherwise considered too costly.
For example, the study in \cite{bao15} leverages
randomized eviction and heterogeneous latencies for a cache architecture.
The authors demonstrate that such techniques incur high performance overheads in 
2D ICs but can be realized even with gains in 3D ICs.

As with side-channel attacks, fault-injection attacks may become more difficult, due to the physical encapsulation of 3D/2.5D ICs.
%% NOTE not included and cited because of Tao Li
%For example, in \cite{4771811}, albeit without such attacks in mind, the authors
%have studied how particle strikes reduce significantly for the inner chips of a 3D IC, 
%resulting in an enhanced reliability and reduced soft error rate for those inner chips.
%
Still, in \cite{rodriguez19} it was shown recently that a lateral re-arrangement of the laser setup can suffice to enable such fault-injection attacks
also for backside-protected 2D ICs and possibly also for 2.5D and 3D ICs.
However, with a dedicated physical design of 3D ICs, e.g., placing TSVs densely at the chip boundaries, forming a ``vertical
shield'' structure~\cite{knechtel19_3D_IOLTS}, along with regular shields in the BEOL and backside protection, even such attacks might remain difficult.

\subsubsection{Data Security at Runtime:
	Physical Read-Out and Probing Attacks}

Similar to fault-injection attacks, the notion of physical enclosures enabled by 3D/2.5D integration may hinder read-out and probing attacks.
	In \cite{knechtel19_3D_IOLTS}, the authors argued for ``all around shields'' enabled by 3D ICs.
Similar protection against probing has been discussed before in~\cite{briais12,cioranesco14}.
While powerful in principle, such schemes are yet to be demonstrated in practice.

\section{Challenges
	%and Prospects
		for Hardware Security Using Emerging Technologies}
\label{sec:outlook}

As with any security scheme, also those relying on emerging technologies face some challenges.
I have outlined specific limitations and challenges for selected prior art already in the discussion above, but here I seek to take a broader view.

Possibly the most important and complex challenges arise from the fact that security schemes in general and those based on emerging technologies
in particular rely on proper technology exploration, device characterization and modeling, circuit architectures, design strategies, etc.
While security schemes tailored for CMOS technology can leverage
well-defined and well-characterized technology libraries, state-of-the-art design tools,\footnote{Even in the context of classical CMOS
	technology, there are still a plethora of challenges to be addressed to make design tools more
		security-aware~\cite{knechtel20_Sec_EDA_DATE}.} etc.,
this is not so straightforward for most emerging technologies, as their characterization and design support is still progressing.
	However, the resulting challenges may---and should---be rather considered as chances,
namely to incorporate security as best as possible right from the beginning, and not only as an
	afterthought, as we often see for security schemes using CMOS technology.

	To do so, I propose that the following steps, among others and not necessarily in that particular order, are to be considered:
	\begin{enumerate}
	\item Establish closer interaction between the communities working on hardware security, physical design, and emerging
		technologies. I argue that a collaborative exchange between our
		communities is essential for any advancement.
	\item (Re-)definition of security metrics and joint ``translation'' of those metrics. Only then can the emerging-technology
	communities consider hardware security in a proper, quantifiable manner during technology exploration and device
	design. For example, while the correlation of switching activities and power consumption (for estimation of power side-channel
			leakage) seems rather easy to model also for emerging technologies,
	the commonly promoted criteria for PUFs, like uniqueness or unpredictability, are highly device-specific and require
	``translation'' for effective modeling.
	\item Joint reconsideration of threat models. With the progression of emerging technologies, it can be expected that further fallacies,
	possibly yet unexplored, come into play. For example, advanced tools for failure analysis of emerging technologies
		can be ``misused'' for physical attacks in the field. For security schemes based on CMOS technology, we have seen this
		with the proliferation and wider availability
		of, e.g., electron microscopy~\cite{sugawara14} and electro-optical probing~\cite{tajik17_CCS}.	
	\item Technology exploration, prototyping, and joint evaluation of securities schemes based on emerging
	technologies. That is also because process variations are expected to be more pronounced for most emerging technologies. For PUFs, e.g.,
	stronger variations would serve well for uniqueness/entropy,
		but hinder reproducibility. Only once such empirical insights are available it would
		seem possible to devise, e.g., error correction schemes to improve the reproducibility.
	\item Since most emerging technologies are being promoted for CMOS integration, it becomes important to determine the ``weakest
	link(s) in the chain'' for such hybrid implementations of security schemes.
	These efforts should range from device to circuit to system level, and should also revise related design strategies as needed.
	Somewhat related, note that the composition of security schemes remains a challenge even for CMOS-only ICs;
	the physical-design community should become more involved here as well~\cite{knechtel20_Sec_EDA_DATE}.
	\end{enumerate}

\section{Conclusion}
\label{sec:conclusion}

In this paper, I have reviewed the fundamentals of hardware security in general and discussed selected prior art. I have further
reviewed selected emerging technologies, namely (i)~spintronics, (ii)~memristors, (iii)~carbon nanotubes and related transistors,
	(iv)~nanowires and related transistors, and (v)~3D and 2.5D integration.
I have discussed how these technologies are promising to advance various aspects of hardware security, and I have also reviewed
related early studies and some of their limitations.
Note that I made an effort to include hyperlinks for each and every reference, which should serve well for further literature study of any interested
reader.

%As with emerging technologies in general,
Many of the security schemes based on emerging technologies still require more collaborative efforts and practical validation---which is not to
be taken as pessimistic statement, rather as call to our communities for joint action.

%\newpage

%\bibliographystyle{ACM-Reference-Format}
\bibliographystyle{IEEEtran}
\balance
\bibliography{main,Knechtel,others}

\end{document}